\documentclass[useAMS,usenatbib]{mn2e}

\usepackage{aas_macros,graphicx,multirow,rotate}

\def\src{GRB\,041219A}

\def\sax {Beppo\emph{SAX}}

\def\int {\emph{INTEGRAL}}

\def\swi {\emph{Swift}}

\title[GRB 041219A and its host galaxy]{A detailed spectral study of GRB 041219A and its host galaxy}
\author[D. G\"{o}tz et al.]{D. G\"{o}tz$^{1}$\thanks{E-mail:
diego.gotz@cea.fr}, S. Covino$^{2}$,  R. Hasco\"et$^{3}$, A. Fernandez-Soto$^{4}$, F. Daigne$^{3,5}$, R. Mochkovitch$^{3}$, \newauthor P. Esposito$^{6}$
\\
\smallskip\\
$^{1}$AIM (UMR 7158 CEA/DSM-CNRS-Universit\'e Paris Diderot) Irfu/Service d'Astrophysique, Saclay, F-91191 Gif-sur-Yvette Cedex, France\\
$^{2}$INAF -- Osservatorio Astronomico di Brera,  Via E. Bianchi 46, 23807 Merate (LC), Italy \\
$^{3}$Institut d'Astrophysique de Paris, UMR 7095
Universit\'{e} Pierre et Marie Curie-Paris 6 -- CNRS, 
98 bis boulevard Arago, 75014 Paris, France\\
$^{4}$Instituto de Fisica de Cantabria, CSIC-Universidad Cantabria, Avenida de los Castros s/n, 39005 Santander, Spain\\
$^{5}$Institut Universitaire de France\\
$^{6}$INAF -- Osservatorio Astronomico di Cagliari, localit\`a Poggio dei Pini, strada 54, I-09012 Capoterra, Italy}

\begin{document}

\date{Accepted . Received ; in original form }

\pagerange{\pageref{firstpage}--\pageref{lastpage}} \pubyear{2002}

\maketitle

\label{firstpage}

\begin{abstract}

\src\ is one of the longest and brightest gamma-ray bursts (GRBs) ever observed. It was discovered by the \emph{INTEGRAL} satellite, and thanks to a precursor happening about 300 s before the bulk of the burst, ground based telescopes were able to catch the rarely-observed prompt emission in the optical and in the near infrared bands.

Here we present the detailed analysis of its prompt $\gamma$-ray emission, as observed with IBIS on board \emph{INTEGRAL}, and of the available X-ray afterglow data collected by XRT on board \emph{Swift}. We then present the late-time multi-band near infrared imaging data, collected at the TNG, and the CFHT, that allowed us to identify the host galaxy of the GRB as an under-luminous, irregular galaxy of $\sim 5\times10^{9}M_{\odot}$ at best fit redshift of $z=0.31_{-0.26}^{+0.54}$.

We model the broad-band prompt optical to $\gamma$-ray emission of \src\ within the internal shock model. We were able to reproduce the spectra and light curve invoking the synchrotron emission of relativistic electrons accelerated by a series of propagating shock waves inside a relativistic outflow. On the other hand, it is less easy to simultaneously reproduce the temporal and spectral properties of the infrared data.
\end{abstract}


\section{Introduction}

Gamma-ray bursts (GRBs) were discovered in the late 1960s \citep{klebesadel73}, and 40 years later they are still not fully understood \citep[see e.g.][for a recent review]{gehrels09}. 
They are short lived transients (ms to hundreds of seconds) of soft gamma ray radiation that appear at random directions on the whole sky. A major breakthrough in GRB science has been achieved in the late 1990s, when thanks to the rapid localizations performed by the Italian-Dutch satellite \sax\ the first transient counterparts, the so-called afterglows,  have been discovered at X-ray \citep{costa97}, optical \citep{vanpa97}, and radio wavelengths \citep{frail97}. These and the subsequent measurements, thanks to the \emph{HETE-II}, \int, and \swi~ satellites have made it possible to firmly prove the cosmological nature of these sources, being their redshifts, $z$, distributed in the range ($0.1 \div 8.2$).
Thanks to the distance measurements, the energy and luminosity of GRBs could be determined making them the most powerful explosions
in the Universe, since their isotropically emitted energy $E_{\rm iso}$ spans from 10$^{50}$ to 10$^{54}$ erg \citep{amati07}.
However, this huge amount of energy request can be reduced in the hypothesis that GRBs are collimated sources \citep{rhoads97}, and indeed the detection of some achromatic breaks \citep[e.g.][]{zeh06} in the light curves of GRB afterglows further supports
this interpretation. From the measure of the break times the jet opening angles, $\theta_{\rm jet}$, can be derived \citep{sari99a}, and taking this into account the energy reservoir and its spread can be reduced, clustering around 10$^{51}$ erg \citep{frail01,bloom03}. 
In several cases late bumps in the afterglow light curves have been reported \citep[e.g.][]{bloom99}. These bumps have been interpreted as an emerging supernova component. In a handful of cases the GRB/SN association has been spectroscopically confirmed \citep[see][for a review]{woosley06}. These arguments and the energetic budgets point towards the collapse of a massive ($>$ 30 M$_{\odot}$) star as being at the origin of GRBs, at least for the events lasting longer than $\sim$2 s \citep{woosley93}.

\src~ was detected and localized in real time by the \int~ Burst Alert System \citep[IBAS;][] {ibas}.
The alert was issued when the burst was still on going, and this allowed some robotic telescopes
to detect a prompt optical and infrared flash \citep{vestrand05,blake05} whose position was consistent with the IBAS one \citep{gotz04}. 

Only a handful of bursts have been observed in the visible range during the prompt phase, and, as often in GRB physics, it is impossible to identify a single common behaviour \citep[see][]{yost07a,yost07b}. GRB 990123 was bright  
and showed no clear link between the three measurements made during the prompt phase and the gamma-ray data \citep{akerlof99,briggs99}. The ``naked eye burst'' GRB 080319B was even brighter, and may show some correlation with a delay of a few seconds between the optical and the high energy emission \citep{racusin08}. GRB 050820A \citep{vestrand06} and GRB 041219A were much weaker (both in apparent and absolute magnitudes) and their optical fluxes were apparently correlated to the $\gamma$-rays. In these two latter cases it seems reasonable to believe that  the prompt emission can be explained by the same mechanism at high and low energy \citep{geneta2007}.
Conversely, the two optically bright bursts probably require different physical conditions or/and a different radiative process in the regions emitting the low and high energy photons respectively \citep{sari99b,meszaros99,zou09,hascoet}. 

\src~ is the longest and brightest GRB in the IBAS sample \citep{vianello09}. Indeed the very good quality of the gamma-ray data allowed \citet{gotz09} to measure the 
time-variable high linear polarization of this event in the 200--800 keV energy band, partly confirming previous but less significant results \citep{kalemci07,mcglynn07}. 
Here we present the detailed spectral analysis of the prompt $\gamma$-ray emission of \src, as measured by IBIS, and SPI-ACS on board \int~ \citep{integral}, of its X-ray afterglow, as measured by \swi/XRT, and the identification of its host galaxy, as observed by the TNG, and the CFHT, as well as the modelling of the broad band, optical to $\gamma$-ray, prompt emission of the GRB. 

\section{Observations and Data Reduction}

\subsection{\int\ observations}

IBAS triggered at 01:42:17 UTC (from now on $T_{\rm 0}$) on December 19$^{th}$ 2004 (IBAS Alert \# 2073), on a precursor of
GRB 041219A indicating a 38.2 sigma source at R.A. = 00$^{h}$ 24$^{m}$ 26$^{s}$, Dec. = +62$^{\circ}$ 50$^{\prime}$  06$^{\prime\prime}$ ($l$ = +119.86$^{\circ}$; $b$ = +0.13$^{\circ}$) with a 2.5$^{\prime}$ uncertainty .
We selected the corresponding \int~ pointing (SCWID: 026600780010), and processed the data using the Off-line Scientific Analysis (OSA) software provided by the \int\ Science Data Centre (ISDC, \citealt{courvoisier03}) v8.0. 

We analysed IBIS \citep[\int\ coded mask imager;][]{ibis} data. 
Our analysis is based on data taken with ISGRI \citep{isgri}, the IBIS low energy detector array,
which is made of CdTe crystals, and is working in the 13 keV--1 MeV
energy range, and on PICsIT \citep{picsit} a pixellated CsI detector, working in the 150 keV--10 MeV energy range.
In the ISGRI 20-100 keV image the only detected source is \src, which is localized at 
R.A. = 00$^{h}$ 24$^{m}$ 27$^{s}$, Dec. = +62$^{\circ}$ 50$^{\prime}$  21$^{\prime\prime}$ with a 19$^{\prime\prime}$ uncertainty, consistent with the IBAS and the infrared \citep{blake05} positions.  We used this
position to extract ISGRI spectra in 62 logarithmically spaced bins between 13 keV and 1 MeV. ISGRI light curves
have been obtained by binning ($\Delta$t=1 s) the ISGRI events in three different energy bands (20--40; 40--100; 100--300  keV). The background has been subtracted by fitting a constant value to these light curves using the data of the same pointing before the bursts itself (i.e. before 01:41:00 UTC). 

The source is also clearly detected in the PICsIT single events histograms (252--336 keV band), but the positional uncertainty in this case is larger and we do not report the PICsIT position here. PICsIT histograms are not suited for spectral extraction in this case,
since they are integrated over the entire duration of the pointing ($\sim$1800 s). For spectral and light curves extraction
we used PICsIT spectral timing data. They have no positional information, since they represent the total count rate of the PICsIT camera  in 4 energy bands (158--208; 208--260; 260--364; 364--676 keV). We used them to extract light curves, rebinned at 4 s intervals. These light curves have been background subtracted by evaluating the background
before and after (next pointing) the GRB. The same data have then been used to extract PICsIT spectra  by summing the count rates relative to the GRB in 4 bands and rebinning the PICsIT response matrix in order to
cope with available energy bands.  Due to the
fact that the GRB was in the fully coded field of view, no further corrections were necessary, and the standard ancillary
response file could be used. For the spectral analysis, background spectra have been derived by using the pointing
before and the pointing after the GRB.

For timing purposes only, we also used the data of the Anti-Coincidence System (ACS) of the \int~ spectrometer SPI \citep{spi}. ACS data consist of the total count rate (50 ms time resolution) above $\sim$100 keV measured by the 91 Bismuth
Germanate (BGO) scintillator crystals that surround SPI. 
The crystals are used as the anti-coincidence system of the spectrometer, but are also
capable of detecting high-energy transient events such as
bright GRB and SGR bursts \citep{acs}.  We computed the ACS light curve with a
bin size of 5 s and estimated the background by fitting a constant value to the data of the same pointing excluding the burst itself.

\subsection{\swi/XRT observations}
At the epoch of \src, the automated slewing of the \swi\ narrow-field instruments was not enabled yet: for this reason the X-Ray Telescope (XRT)  did not begin collecting data until about 5 hours after the IBIS detection. \swi/XRT observed the afterglow for roughly 0.6 ks in windowed timing (WT) mode and for about 1 ks in low rate photodiode (LR) mode \citep[see][for more details on the XRT data-taking modes]{hill04}. Given the low count rate, here we only considered WT data.

The data were processed with standard procedures and screening criteria using the \textsc{xrtpipeline} (version 0.12.3 under the \textsc{Heasoft} package 6.6).   We extracted the source events from a $40\times40$ pixels (1 pixel corresponds to about 2.36 arc sec) box along the image strip; a region of the same shape and size, positioned well outside the point-spread function of the source, was used to estimate the background. The ancillary response file was generated with \texttt{xrtmkarf}, and it accounts for vignetting and point-spread function corrections.

\subsection{TNG/NICS observations}
The first attempt to identify the host galaxy of GRB 041219A, was made through an H-band observation of the field of the GRB with the NICS camera mounted on the 3.5 m TNG\footnote{http://www.tng.iac.es/} telescope at La Palma. 

Observations of the field of \src\ were performed on Oct 15$^{th}$ 2007, i.e. a few years after the GRB. The observations were secured under good observing conditions with about 1\,\arcsec\ seeing in the $H$ band. In total 15 frames with 30\,s exposure and NDIT=3 were collected. Data reduction was carried out following standard procedures with the Eclipse package \citep{Dev01}. The median of the input frames was derived to obtain a sky frame which was subsequently subtracted from the input frames. Then sub-pixel registration was applied, and a final scientific frame was obtained averaging all input frames. An astrometric solution was derived by a large number of 2MASS\footnote{http://www.ipac.caltech.edu/2mass/} objects and the final absolute accuracy is about 0.5\arcsec\ both in R.A. and Dec. Aperture photometry was computed with tools provided by the GAIA package\footnote{http://star-www.dur.ac.uk/$\sim$pdraper/gaia/gaia.html} and the absolute calibration was derived by comparison with a suitable number of not-saturated, isolated 2MASS stars in the field.

\subsection{CFHT/WIRCam observations}
Further observations of the field of GRB 041219A were obtained during the night of October 6$^{th}$ 2009 (program ID: 09BF08) with the WIRCam instrument at the 3.6 m CFHT in Mauna Kea under good sky conditions (seeing 0.5--0.8$^{\prime\prime}$), providing a complete set of photometric images. The region of \src\ was observed in the $Y$, $J$, $H$, and $K_{\rm s}$ filters for about 30 minutes per filter.

In order to perform sky subtraction, a first stack of the images has been produced using a median of the input images.  Saturation masks have been generated  looking for saturated pixels in the individual images, in order to discard saturated objects from the catalogues. A global astrometric solution has been found for the whole dataset using the 2MASS catalogue as an absolute reference. A single binary mask, masking all the sources detected with {\tt sextractor} has then been produced, and the sky value for each pixel could be determined.
On the sky subtracted images relative photometry and astrometry could then be performed.
The internal dispersion (difference between same objects coordinates in different images) of the astrometric solution for high S/N objects is small, $\sim$0.06$^{\prime\prime}$, and the reference dispersion (difference between the object coordinates in our images and the 2MASS catalog ones) of the astrometric solution for the high S/N objects is around 0.13$^{\prime\prime}$ (similar to the internal accuracy of the 2MASS astrometry).

The stability of the sky emission was not perfect during this run. As a consequence, a number of images show strong large scale gradients in the background, even after the sky-subtraction, because of the short time scale fluctuations of sky emission lines. The following procedure was applied to remove this pattern before the combination of the images: we ran {\tt sextractor} on the individual images with a mesh size of 256 pixels and using a very aggressive masking of the objects (produced during the sky-subtraction step). The output background image was then subtracted to the input image. 

One stack is finally produced for each of the four filters ($Y$, $J$, $H$, and $K_{\rm s}$). 
The absolute photometry is then checked: while for the $J$, $H$, and $K_{\rm s}$ bands the 2MASS catalog has been used as a photometric reference, for the $Y$ band zero points have been adopted from CFHT observations of standard stars\footnote{ http://www.cfht.hawaii.edu/Instruments/Imaging/\\WIRCam/IiwiVersion1Doc.html\#Part6}.

\section{Analysis and Results}
\label{sec:results}
\subsection{{\int}}
\subsubsection{Timing analysis}
\label{integral:timing}

\src, being the brightest and longest burst detected by \int, is the only one for which good PICsIT data are available. Therefore we included them in our analysis. In this section we present the analysis of this GRB using both IBIS detector planes and the Anti-Coincidence System (ACS) of the \int\ spectrometer.

\begin{figure*}
\resizebox{\hsize}{!}{\includegraphics[angle=90]{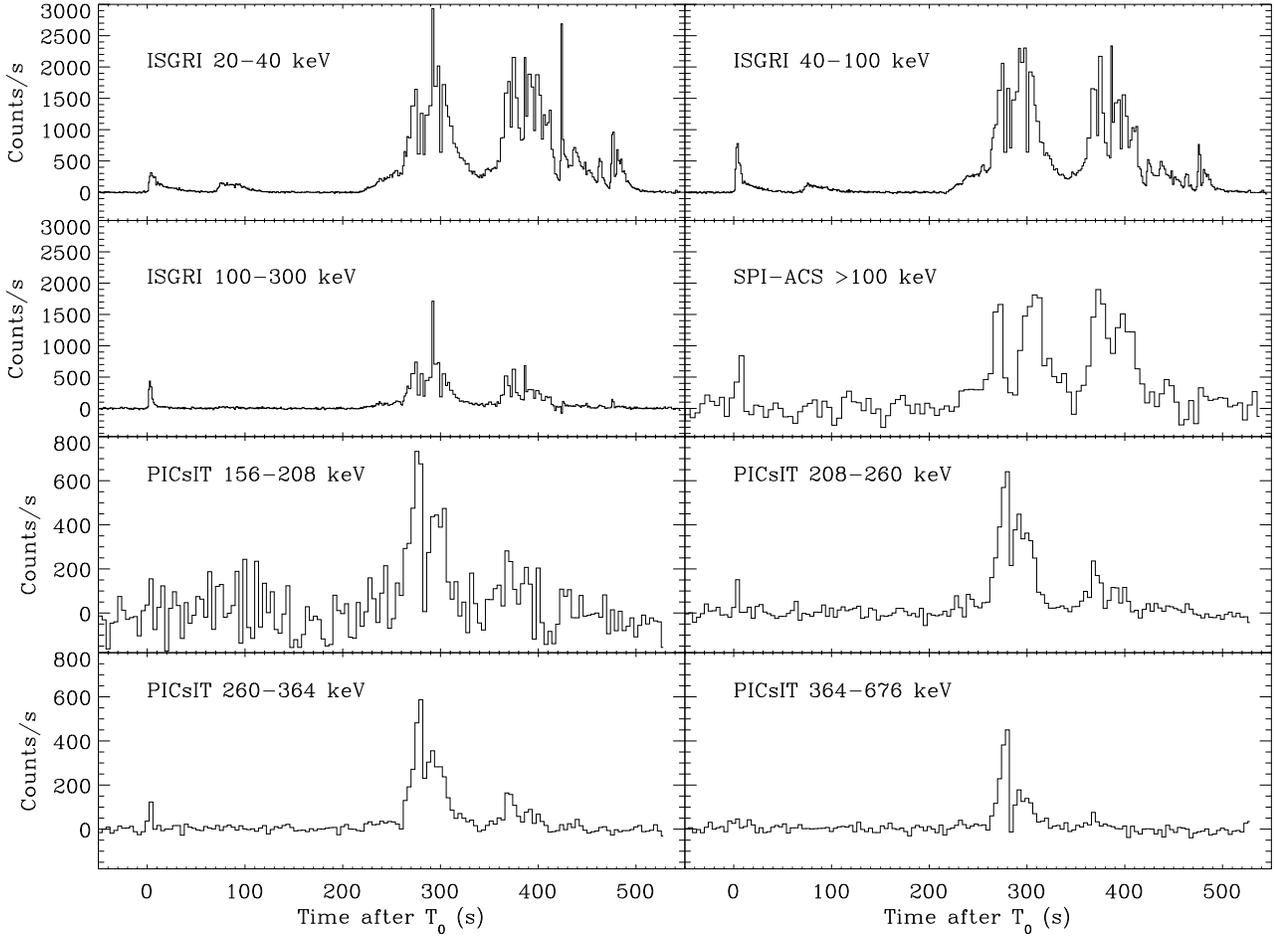}}
\caption{\label{fig:lc}Background-subtracted light curve of GRB 041219A as seen by ISGRI, PICsIT, and the Anti-Coincidence System of SPI in different energy bands. The error bars have been omitted for clarity.}
\end{figure*}

The light curve of \src, as measured by the different high-energy instruments is shown in Fig. \ref{fig:lc}. The GRB
consists of two main peaks, which are preceded by two precursors, the second of which is detected only in
the ISGRI data. The ISGRI data are affected by telemetry saturation for what concerns the first precursor, and the
two main peaks, hence these light curves are not representative of the true ISGRI recorded flux. PICsIT and ACS data,
also shown in Fig. \ref{fig:lc}, are complementary to the ISGRI ones 
in the sense that they extend the observable energy range to higher values, but, in view of the globally smaller number of registered counts, they are not telemetry limited on what concerns the two main peaks.

\subsubsection{Spectral analysis}
\label{integral:spectra}

ISGRI spectra have been derived for the four different sections of the GRB identified as first and second precursor, and first and second peak, as defined in Table \ref{tab:int}. 
PICsIT spectra were obtained by integrating the spectral timing data over the same time intervals considered
for the ISGRI analysis. Only for the first and second peak PICsIT data have enough statistics to be included in the spectral fits. 

The data were grouped with a minimum of 20 counts per energy bin and the spectra were analysed with the \textsc{xspec} fitting package version 11.3.
A multiplicative factor has been added to the joint fits in order to take into account the instruments inter-calibration. The two
datasets never needed to be adjusted by more than 30\%.
For fitting purposes a simple power law model has been considered
as a first approach, then an exponential high-energy cut-off has been adopted to better describe the data, and finally
a ``Band" GRB model has been used if necessary. 
The results of the fits are reported in Table \ref{tab:int}. Due to the high flux, 3\% of systematics has been added to the
spectra during the fit.

\begin{figure}
\centering
\resizebox{\hsize}{!}{\includegraphics[width=5.5cm, angle=-90]{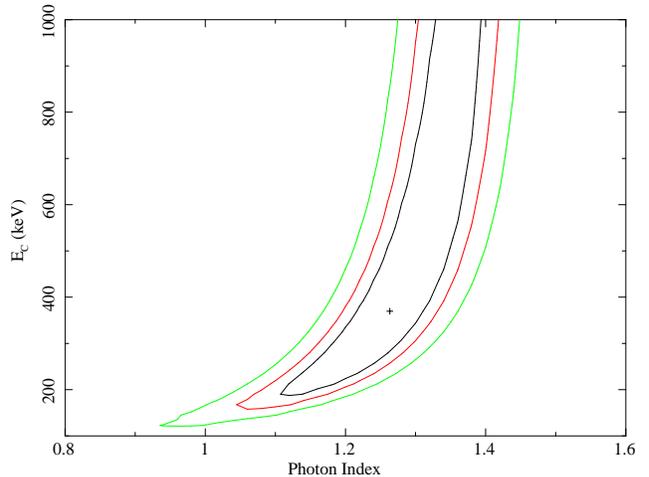} }
\caption{Confidence contours (68\%, 90\% and 99\% c.l.) of the cut-off energy $E_{c}$ versus the photon index $\Gamma$ for the ISGRI fit of the first precursor.}
\label{fig:cc}
\end{figure}

The first precursor has a very hard spectrum with respect to the rest of the GRB, as is seen also from the light curves inspection (see Fig. \ref{fig:lc}).
It can be well fit by a single power law model, but a slight improvement can be achieved if the spectrum is
fitted with a cut-off power law model, see Table \ref{tab:int}. Even if the cut-off energy, $E_{\rm c}$, is not well constrained for what concerns
its maximal value, a solid lower limit can be derived, as shown by the contour plot (photon index vs. cut-off energy) of Fig. \ref{fig:cc}.
The derived (90\%) lower limit for the cut-off energy ($E_{c}$) is 180 keV , corresponding to an $E_{\rm peak}$ energy\footnote{$E_{\rm peak}$ is given by (2-$\Gamma$)$\times E_{\rm c}$ in this case.} larger than 130 keV, and a best fit value of 370 keV, making this part the hardest of the whole GRB.

The second precursor shows a soft spectrum, and it can be described by a simple power law model without the need of any cut-off. Other models like a black body or the quasi-thermal model (see below) can be ruled out ($\chi^{2}$/d.o.f. = 343.5/46, and $\chi^{2}$/d.o.f. = 57.3/42, respectively). It is the softest part of the whole GRB, as it is not detected above 100 keV. 

\begin{figure}
\centering
\resizebox{\hsize}{!}{\includegraphics[width=5.5cm, angle=-90]{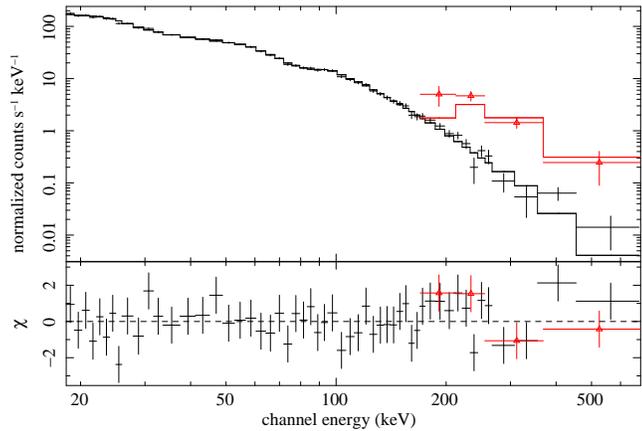}}
\caption{Simultaneous ISGRI/PICsIT spectral fit of the first peak of \src. In the upper panel the data points (crosses for ISGRI and triangles for PICsIT) are compared to the best fit model, while the lower panel represents the residuals with respect to the fit.}
\label{fig:sp1nu}
\end{figure}

The first main peak is very bright, and in this case the single power law fit can be excluded ($\chi^{2}$/d.o.f. = 116.63/59), while the cut-off power law model and the Band model give an equally good representation, with 
$\chi^{2}$/d.o.f. = 72.29/58 and $\chi^{2}$/d.o.f. = 72.11/57. But having the Band model an extra degree of freedom, and being the parameter $\beta$ not well constrained, 
in Table \ref{tab:int} we report only the fits done using the cut-off power law model.
The simultaneous ISGRI and PICsIT best fit model and data are shown in Fig. \ref{fig:sp1nu}.

The second peak is fainter and softer than the first one. As for the first peak, it can equally well be fit by a Band model or by a cut-off power law, and the best fit of the joint ISGRI/PICsIT data is reported in Table \ref{tab:int}. 
The simple power law model, on the other hand, is excluded ($\chi^{2}$/d.o.f. = 103.97/54). The derived
$E_{\rm peak}$ in this case is significantly smaller than the one derived for the first peak, as can be seen by comparing the two values in Table \ref{tab:int}.


\begin{table*}
\centering
\caption{\int\ spectral parameters. Errors are given at 90\% c.l.} 
\label{tab:int}
\begin{tabular}{ccccccccc}
\hline
Name & T$_{\rm Start}$ & T$_{\rm Stop}$ & $\Gamma$ & $E_{\rm c}$ & $E_{\rm  peak}$&$\chi^{2}$&d.o.f&PICsIT\\ 
          & UTC           & UTC           &                    & keV       &    keV        &                & &\\ 
\hline
First Precursor & 01:42:17 & 01:43:12 & 1.41$\pm$0.04 &-&-&60.52 &49&no\\ 
  & 01:42:17 & 01:43:12 & 1.26$\pm$0.13 & $>$180$^{a}$& $>$134$^{a}$&57.67 &48&no\\ 

Second Precursor &  01:43:27 & 01:44:12 & 2.16$\pm$0.06&-&-&31.41 &44&no\\
First Peak &  01:46:22 & 01:47:40 &1.42$\pm$0.05&347$_{-72}^{+138}$&201$_{-41}^{+80}$&68.37&58&yes\\ 
Second Peak &  01:48:12 & 01:48:52 & 1.71$\pm$0.06&334$_{-97}^{+195}$&97$_{-28}^{+56}$& 78.03&52&yes \\ 
1$^{st}$ peak+ 2$^{nd}$ peak &  01:46:02 & 01:50:42 & 1.53$\pm$0.01&353$_{-50}^{+171}$&166$_{-23}^{+81}$&77.59 &62 &yes\\ 
\hline 
\end{tabular}
\begin{list}{}{}
\item[$^{a}$] The best fit values are 370 keV and 274 keV for $E_{\rm c}$ and $E_{\rm peak}$, respectively.
\end{list}
\end{table*}

The average ISGRI/PICsIT spectrum of the two main peaks is also well fit by a cut-off power law, and its parameters are also reported in Table \ref{tab:int}. Its fluence in the 20--200 keV energy band is 2.5$\times$10$^{-4}$ erg cm$^{-2}$. The bolometric (1 keV--10 MeV) fluence of the whole GRB is $\sim 5 \times$10$^{-4}$ erg cm$^{-2}$.
The derived spectral parameters can be used to determine the $pseudo-z$ \citep{atteia03,pelangeon06} for \src, which results to be 0.19$\pm$0.06. Using this distance estimator the isotropic equivalent energy released by this GRB would be of the order of 2$\times$10$^{52}$ erg. 

We note that, except for the first precursor, the spectral parameters derived here are compatible within the errors with the ones derived
by \citet{mcbreen06}, obtained using the germanium spectrometer, SPI, on board \int. The differences for
the first precursor, that in our analysis is softer, may result from the much better statistics we have at low energies 
in the current analysis. Furthermore using the quasi-thermal model, composed by the sum of a black body and a power law, for modelling the rest of the GRB, as proposed by the same authors, we cannot confirm their findings in terms of black body temperatures, obtaining $kT=16\pm6$ keV for the first peak and $kT=19\pm5$ keV for the second, while they obtain much higher values.

\begin{figure}
\centering
\resizebox{\hsize}{!}{\includegraphics[width=9cm,angle=90]{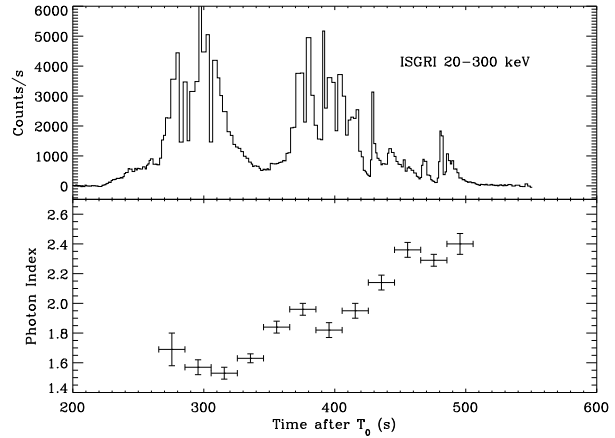} }
\caption{Time resolved spectral analysis of the IBIS/ISGRI data. The derived photon index values are plotted against time (lower panel), and the 20-300 keV light curve is plotted in the upper panel, as a reference.}
\label{fig:timeres}
\end{figure}

Thanks to the fact that \src\ is very bright, we have been able to perform a detailed time-resolved analysis. We extracted contiguous spectra integrated over 20 s during the main GRB emission (i.e. excluding the precursors). Due to the lower signal-to-noise ratio, the spectra were just fitted with simple power-law models. The results are presented in Figure \ref{fig:timeres}. As can be seen, a classical hard-to-soft evolution is observed, and we note that the photon index $\Gamma$ evolves from about 1.5 to about 2.5, indicating that while $E_{\rm peak}$ decreases during the burst (as indicated by the values in Table \ref{tab:int}) it literally crosses the IBIS/ISGRI energy domain, since the values of $\Gamma$ measured at the end of the burst are likely to be the asymptotic value of the $\beta$ parameter of the Band function.

\subsection{X-ray afterglow}

The observed net count rate during the WT mode observations was as low as $0.60\pm0.03$ counts s$^{-1}$ between 0.3 and 10 keV, so no pile-up correction was necessary. The data were grouped with a minimum of 20 counts per energy bin. 
Given the paucity of counts, we fit a simple model to the data: a power law corrected for absorption. We obtained the following best-fitting ($\chi^2_{\nu}=1.0$ for 14 degrees of freedom) parameters: absorbing column $(1.5\pm0.3)\times10^{22}$ cm$^{-2}$ and photon index $\Gamma=1.9\pm0.3$.
The observed (not corrected for absorption) 0.3--10 keV flux was $\sim 4.1^{+0.2}_{-0.6}\times10^{-11}$ erg cm$^{-2}$ s$^{-1}$ at $T_{\rm 0}$ + 5 hours, while
the absorption corrected one was $\sim 6.9^{+0.3}_{-1.0}\times10^{-11}$ erg cm$^{-2}$ s$^{-1}$.
 All uncertainties are quoted at 1$\sigma$ confidence level.
We note that the absorbing column density we derive is larger than the value reported by \citet{kalberla05} ($\sim$8$\times10^{21}$ cm$^{-2}$), but the two values are  compatible at a $\sim$2 $\sigma$ level, and the resolution of the LAB HI survey, being of the order of 0.6$^{\circ}$, is too coarse for clumpy low-latitude Galactic regions. But the difference may also be accounted for by the absorption within the GRB host galaxy, being this excess value typical of GRB afterglows \citep[e.g.][]{campana10}.

\subsection{Near Infrared Photometry}

We have observed the field of \src, obtaining first $H$-band
photometry with the NICS camera at the Italian Telescopio Nazionale
Galileo, as well as later photometry in the $YJHK_{\rm s}$ bands with WIRCam
mounted at the Canada-France Hawaii Telescope. One object is
clearly detected in our images with a position compatible with the one
of the prompt infrared flash \citep{blake05}.

\begin{figure}
\centering
\resizebox{\hsize}{!}{\includegraphics[width=9cm,angle=-90]{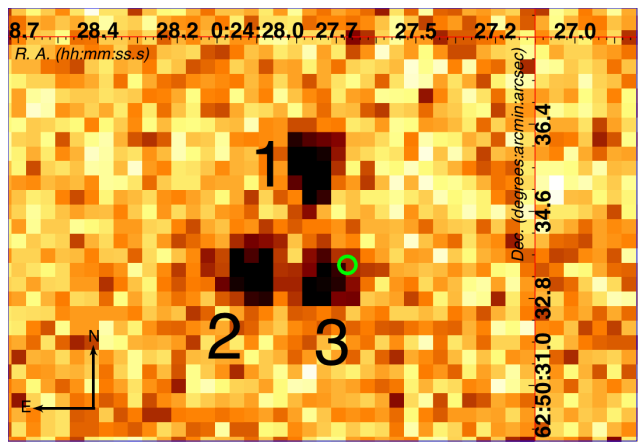} }
\caption{CFHT/WIRCAM $K_{\rm s}$-band image of the region around \src. The circle represent the position of the GRB infrared flash, as derived by \citet{blake05}. The black labels correspond to the three objects discussed in the text.}
\label{fig:CFHT}
\end{figure}

Data corresponding to this object (labeled ``Object 3", see Fig. \ref{fig:CFHT})
are listed in Table \ref{obj3data}. We only list the CFHT results,
whose higher quality supersedes the TNG data. We must remark that the
position of this source is within the Galactic plane ($|b| < 1^\circ$), and
thus it is expected to be heavily reddened.

\begin{table}
  $$
         \begin{array}{p{0.5\linewidth}l}
            \hline
            \noalign{\smallskip}
            Parameter      &  Value \\
            \noalign{\smallskip}
            \hline
            \noalign{\smallskip}
            R.A. (J2000)      & 00^{h}24^{m}27.6^{s}  \\
            Dec. (J2000)       & +62^{\circ}50^{\prime}33.5^{\prime\prime}  \\
            $Y$  &  22.16 \pm 0.35 \\
            $J$  &  20.81 \pm 0.20 \\
            $H$  &  19.73 \pm 0.17 \\
            $K_{\rm s}$  &  18.86 \pm 0.12 \\
            \noalign{\smallskip}
            \hline
         \end{array}
  $$
  \caption[]{Data corresponding to the GRB host, as obtained through
  our CFHT observations. All magnitudes are given in the Vega-based 
  Johnson system, and are not corrected for Galactic extinction.}
         \label{obj3data}
\end{table}

\subsubsection{Object Identification}
\label{sec:identification}

\begin{figure}
\centering
\includegraphics[angle=0,width=9cm]{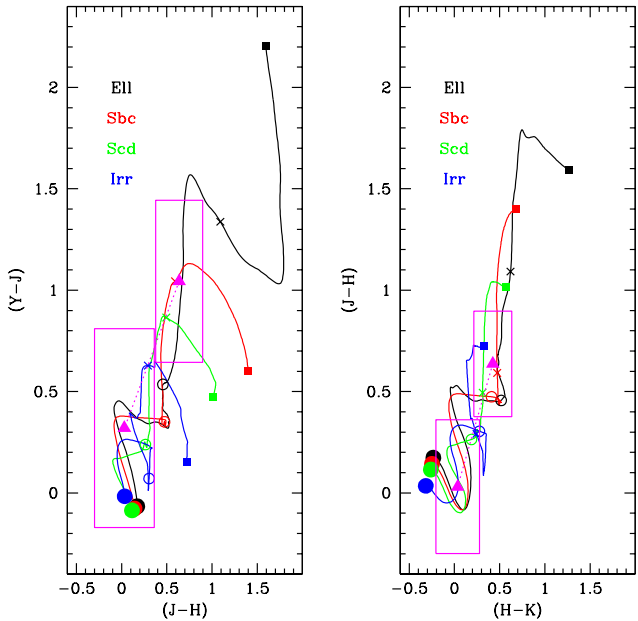}
\caption{Four-band $YJHK$ colour-colour diagrams for the afterglow host. In both panels we present the AB-magnitude colour tracks for different galaxy templates, ranging from elliptical to irregular as indicated in the legend therein. Each track begins at redshift $z=0$, in the position marked with a filled circle, close to colour = 0. From there the redshift increases, passing through $z=1$ (empty circle), $z=2$ (cross), and reaching $z=3$ (filled square). The observed and de-reddened colours of the GRB host are marked with triangles, surrounded by boxes which indicate the associated 1-sigma errors. We have added in quadrature an extra error in the de-reddened colours to include the uncertainty in the reddening value, assuming a 10\% uncertainty in each term.  }
      \label{figcolours}
\end{figure}

We have compared the $YJHK_{\rm s}$ colours of Object 3 to several models of
different galaxy types, in order to obtain some information about the
probable redshift and spectral type. In Fig.
\ref{figcolours} we present the results of such comparison.
We have performed the same analysis for two nearby objects (labeled 1 and 2, and already noticed by \citealt{blake05}), but they were not compatible with typical colours of galaxies of any type.  The
objects have been de-reddened using the local value as given in
\citet{schlegel98}, which yields $E(B-V) = 1.858$ and $A_V =
5.760$ mags -- the correction is also shown in the plots. Given the low Galactic latitude
of the GRB and the coarse resolution of the Schegel maps (2$^{\circ}$), we checked the
absorption value against our X-ray afterglow observations applying the relation between
optical extinction and hydrogen column density derived by \citet{guver09}, $N_{\rm H}$ (cm$^{-2}$) 
$= (2.21 \pm 0.09)\times 10^{21} A_{\rm V}$ (mag). Using the $N_{\rm H}$ value derived from our
X-ray afterglow observations, we obtain $A_{\rm V} = 6.8 \pm 1.6$ mags, which is compatible within the errors with the value derived by \citet{schlegel98}, that has hence been used as the reference value in this work.

Figure \ref{figcolours} shows that the de-reddened colours of Object 3
are compatible with moderate redshift solutions. We have indeed used the
photometric redshift code by \citet{fernandez99} to estimate
the redshift and basic properties of the host galaxy, finding that the
best fit solution corresponds to an irregular galaxy template at
redshift $z=0.31_{-0.26}^{+0.54}$. We must remark, however, that there is a high degeneracy of the models in that region of colour space, which renders possible a large subset of galaxy models, spanning from early-type galaxies at low redshifts ($z\sim 0.1$ to $z \sim 0.5$) to late-type models at moderate redshifts ($z\sim 0.2$ to $z \sim 1.8$).
Indeed by taking all the possible galaxy types into account the best fit redshift is $z=0.3$, with a bimodal 1$\sigma$ confidence interval in the ranges 0.1--1.1 and 1.3--1.7 (secondary minimum). The uncertainties are estimated by evaluating the observed photometric uncertainties, that are included in the fitting procedure, and the systematic ones, induced by the inability of the spectral templates to include all the variability present in galaxy spectra. \citet{fernandez02}, using $BVRIJHK$ data, showed that the latter could be modelled with a normal error distribution with a variable sigma $\sigma(z)/(1+z) \approx 0.065$. Given our similar set-up, but the different wavelength coverage, we estimate that $\sigma(z)/(1+z) \approx 0.10$ is an adequate assessment of the systematic error. Using this estimate and convolving the two error terms as in \citet{fernandez02} we calculated the confidence interval mentioned above.

However, there are other pieces of information that help in solving this
degeneracy. In particular, an application of the Amati relation 
\citep{amati02} clearly prefers the redshift range $z < 0.5$, and the
pseudo-redshift technique (see \ref{integral:spectra}) yields a value $z = 0.19 \pm
0.06$\footnote{Those two indicators are not independent, as the
Amati relation is one of the factors taken into account in the
calculation of the pseudo-redshift as described by \citet{atteia03}.}. In addition, regarding the spectral properties, an early-type galaxy is, according to
previous experience \citep[e.g.][]{savaglio09}, less likely to be the host of a long
GRB.

\begin{figure}
\centering
\includegraphics[angle=-90,width=9cm]{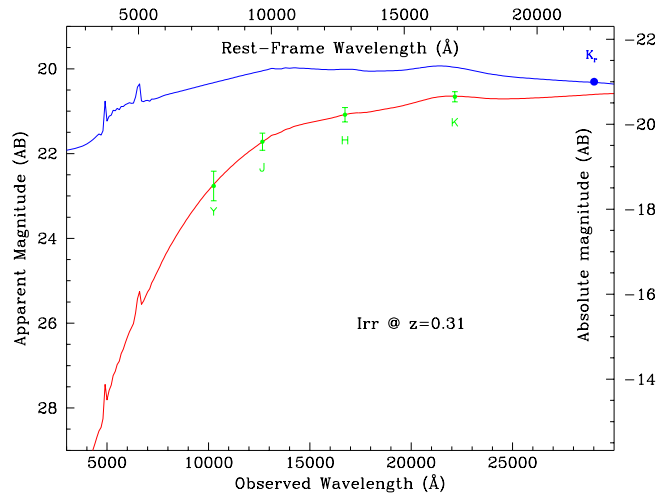}
\caption{Observed photometry of the \src\ host galaxy, together
  with our best-fit spectrum, corresponding to an irregular galaxy at
  redshift $z=0.31$. We also include in the plot the de-reddened
  spectrum (upper line), corresponding to an intrinsic, rest-frame
  absolute magnitude $K=-21\ AB$.}
      \label{fig:spec}
\end{figure}

Taking all this information into account, we have adopted the best-fit
$z=0.31$ as reference value for the redshift, and an irregular
template as the spectral choice. With these values, we show in Fig.
\ref{fig:spec} the intrinsic and observed spectrum of such an
object. As can be seen, the photometry fits the observations
perfectly, and we can derive an absolute rest-frame $K$-band magnitude
$K=-21.0\ AB$, which corresponds approximately to a $\sim 0.1\ L^*$ \citep{fontana06}
galaxy, and, using a simple relation \citep[e.g.][]{savaglio09}, to a mass of $\sim 5 \times 10^9 M_{\odot}$. The derived redshift yields a luminosity distance of 1.6 Gpc
(for $H_{\rm 0}$ = 71 km s$^{-1}$ Mpc$^{-1}$) and a 
total isotropic emitted energy $E_{\rm iso}\sim 10^{53}$ erg (1 keV--10 MeV).
However, taking the uncertainty on the photometric redshift into account the allowed intervals for the quantities cited above are $[-16.5,\ -23.6]$ for the $K\ AB$ magnitude, 
$[3\times 10^{7},\ 7\times 10^{10}]\ M_{\odot}$ for the mass, and 
$[2\times 10^{51},\ 2\times 10^{54}]$ erg for $E_{\rm iso}$. 

\subsection{Broad-band SED modelling of the prompt emission}
\subsubsection{Constructing the broad-band SED}
Thanks to the rapid response to the IBAS alert of the RAPTOR telescope,
\citet{vestrand05} were able to measure the optical emission
of \src\ during its prompt phase. There are three R-band data points and
two upper limits that indicate that the optical emission is variable and likely correlated to the $\gamma$-ray one (see Fig. 3 of their paper). 
\citet{blake05} recorded
the emission of \src\ in three infrared bands, but just on the very final part of the prompt $\gamma$-ray emission. We use these visible and 
infrared data in conjunction to those at high energy presented in this work to build the broad-band SED of the prompt GRB emission at different times.


To do this we extracted four $\gamma$-ray spectra
corresponding to the three time periods where the optical emission has been recorded, and to the one where the infrared data are available. 
In Table \ref{tab:sed} we report the selected time intervals we used and the results of the fits. The infrared data have been de-reddened using 
the same parameters as in Section \ref{sec:identification}.

\begin{table*}
\caption{Derived $\gamma$-ray spectral parameters, optical and infrared fluxes for the four intervals with simultaneous optical and infrared data.} 
\label{tab:sed}
\centering
\begin{tabular}{ccccccc}
\hline 
Selected Time  & Time since $T_{\rm 0}$ & $\Gamma$ & $E_{\rm c}$ & $E_{\rm Peak}$& $\chi^{2}$/d.o.f&Flux density$^{1}$ (filter)\\ 
Interval (UTC) &  s &      & keV     & keV   & & mJy              \\ 
\hline
01:45:43--01:46:56& 211 -- 284 & 1.3$\pm$0.1 & 251$_{-74}^{+198}$& 175$_{-51}^{+139}$ &50.6/56& 2.9$\pm$0.8 ($R_{\rm C}$)\\
01:47:08--01:47:38& 296 -- 326 &1.44$\pm$0.08& 336$_{-101}^{+212}$& 188$_{-56}^{+118}$ &50.6/56& 10.3$\pm$1.1 ($R_{\rm C}$)\\ 
01:47:51--01:49:04& 339 -- 412 &1.79$\pm$0.07 & 401$_{-133}^{+311}$& 84$_{-53}^{+58}$ &67.0/56& 3.8$\pm$0.8 ($R_{\rm C}$)\\ 
01:49:16--01:51:54& 424 -- 582 &2.41$\pm$0.03 & -- & -- & 56.9/46& $<$1.1 ($R_{\rm C}$); 1.7$\pm$0.1 (J); 2.0$\pm$0.1 (H); 2.6$\pm$0.1 ($K_{\rm s}$)\\
\hline 
\end{tabular}
$^{1}$Corrected for Galactic absorption (see text). $R_{\rm C}$ data points are taken from \citet{vestrand05}, while the infrared data points are taken from \citet{blake05}.
\end{table*}

For the first three intervals a cut-off power law model was needed to fit the $\gamma$-ray data, while for the last one no cut-off could be measured, and as stated before, we  may be measuring the $\beta$ parameter of a Band function (see Sec. \ref{integral:spectra}).

\subsubsection{Modelling the prompt emission of GRB 041219A}

\begin{figure}
\begin{tabular}{c}
\includegraphics[scale=0.32]{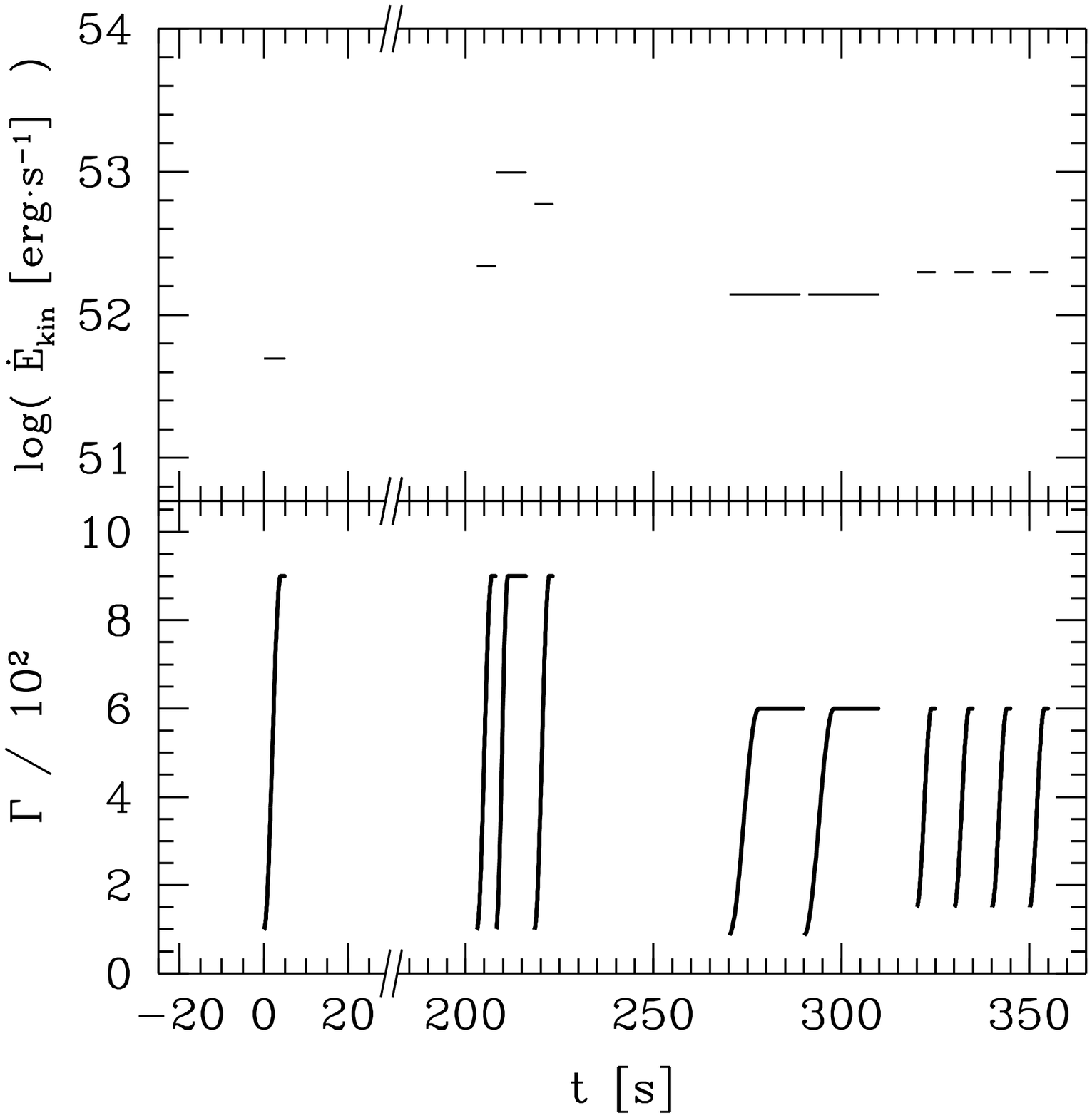} \\
\includegraphics[scale=0.32]{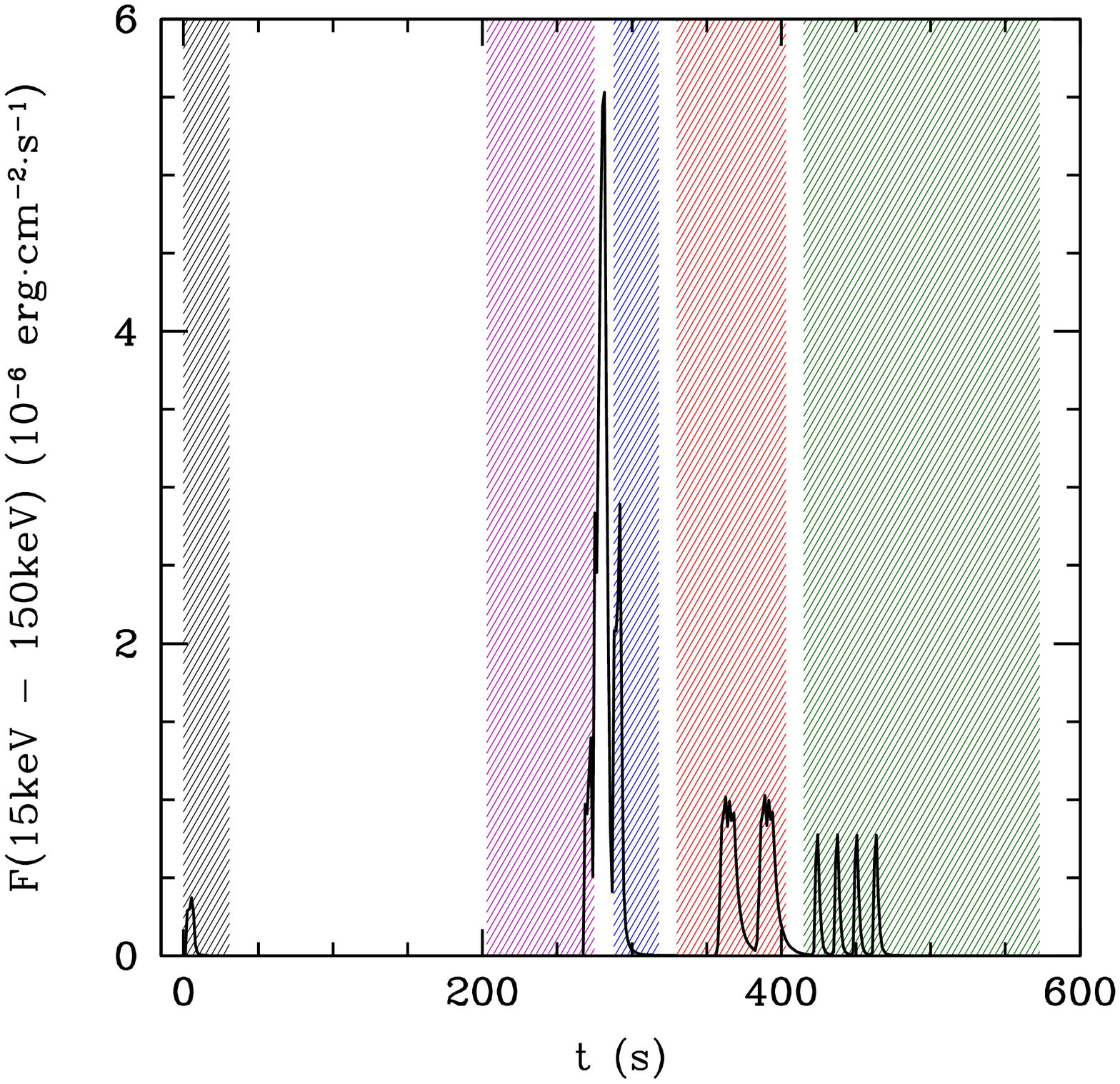} \\
\includegraphics[scale=0.32]{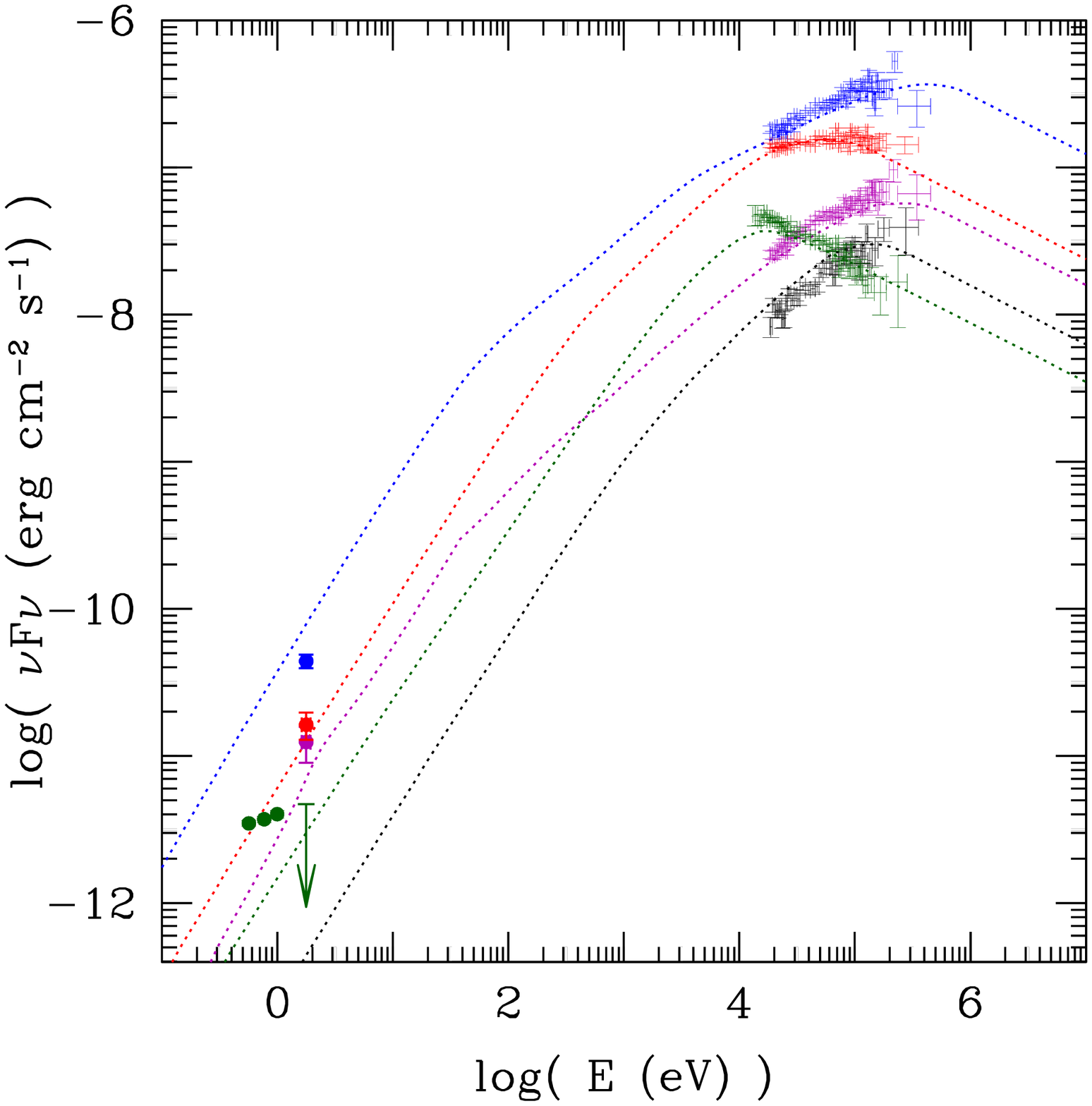}
\end{tabular}
\caption{Model for the prompt emission of GRB 041219A. The top panel shows the adopted initial distribution of the Lorentz factor in the 
flow together with the value of the kinetic power for each injection episode as a function of the injection time $\mathrm{t_{inj}}$. The middle panel represents the corresponding
synthetic light curve in the 15--150 keV interval as a function of observer time t after $T_{\rm 0}$. The five coloured areas represent the integration times for the spectra in the lower panel. 
Except for the first interval they also correspond to the integration times of the optical 
measurements. The lower panel compares the five synthetic spectra to the multi wavelength data. The $JHK$ fluxes 
were obtained during
the last time interval where the burst was not detected in the optical band.}
\label{fig:bb1}
\end{figure}

To model the SED of the prompt phase of GRB 041219A from the infrared to the $\gamma$-ray range we have 
considered the standard scenario of internal shocks \citep{rees94} assuming that the observed 
radiation comes from the synchrotron 
emission of shock accelerated electrons \citep{piran99, bosnjak09}. In order to reproduce the complex time history of the burst 
we have supposed that the initial Lorentz factor distribution in the relativistic outflow 
is made of a succession of elementary injections.
The kinetic power is not the same for all injection phases but remains constant during one given episode (see top panel of Fig.~\ref{fig:bb1}).
The total injected kinetic energy is ${E}_{\rm kin} = 1.95\times 10^{54}$ erg. The adopted micro-physical parameters, which
are necessary to estimate typical values for the post-shock electron Lorentz factor and local magnetic field,
are : $\epsilon_e = 0.33$, $\epsilon_B = 0.0033$ and $\zeta= 10^{-3}$ (fraction of electrons which are accelerated).
The spectrum for each individual shock is computed using the standard synchrotron prescriptions of \citet{sari98}.

Each elementary injection episode develops its own system of internal shocks and a fraction $\epsilon_e$ 
of the dissipated energy is then radiated by the electrons resulting in a  
simulated gamma-ray light-curve represented in the middle panel of Fig.~\ref{fig:bb1} \citep[see][for a detailed description of the model]{daigne98}.
We do not try to get an exact fit of the observed light curve, but it can be seen 
in Fig.~\ref{fig:bb1} that the different phases and corresponding levels of the prompt 
activity are correctly reproduced. 

The simulated spectra for the first three time intervals are also in reasonable agreement with the observations (Fig.~\ref{fig:bb1}, bottom panel)
while the infrared data during the final interval appear more difficult to reproduce (see Sect.~\ref{Infrared emission} below). The adopted slope 
for the population of accelerated electrons is $p=2.8$ in order to get  the correct photon spectral index in the fourth interval, assuming
that it indeed represents the $\beta$ parameter of a Band function. In the other intervals we have no way to constrain
$\beta$. We have then kept the same value $p=2.8$.       
The maximum of each spectrum (in $\nu F_{\nu}$) is located at the synchrotron frequency $\nu_m$ while 
the evolving cooling frequency $\nu_c$ remains located between $\nu_m$ and the optical domain. This
allows us to reproduce 
the different optical levels in the first three time intervals. Also note that the self-absorption frequency $\nu_a$
always remains well below the optical band.


\subsubsection{Deceleration by the external medium}

The discussion presented in the last section only takes into account internal shocks and neglects the deceleration effect by
the burst circumstellar medium. This is justified as long as the typical radius of internal shocks is smaller than the deceleration radius.
In the case of GRB 041219A, which had a very long duration and varied over a wide range of time scales (from less than one second 
to several tens of seconds), this condition implies that the burst environment must have a low density : $n< 0.1$ cm$^{-3}$ for a uniform medium
and $A_*<0.01$ for a stellar wind where $A_*$ is the wind parameter $A={\dot M}/4\pi v$ normalized to $5 \times 10^{11}$ g cm$^{-1}$ 
which is the value expected for a standard Wolf-Rayet wind having ${\dot M}=10^{-5}$ M$_\odot$ yr$^{-1}$ and 
$v=1000$ km s$^{-1}$ \citep{chevalier99,eldridge06}. These limits are quite restrictive, and, if the density is larger, the reverse shock will propagate inside the jet 
before the end of the prompt phase, therefore affecting the dynamics.

\begin{figure}
\begin{tabular}{c}
\includegraphics[scale=0.32]{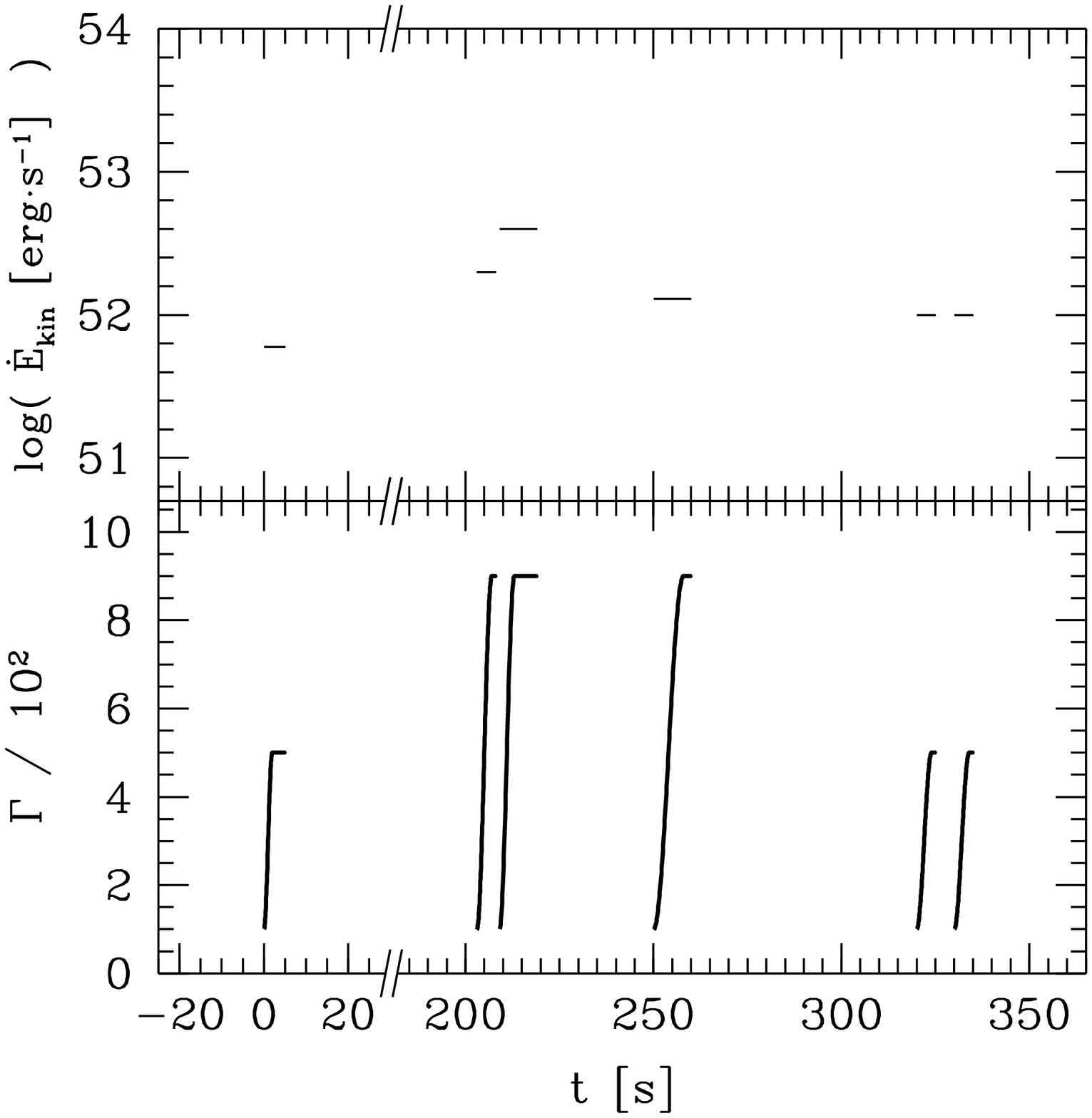} \\
\includegraphics[scale=0.32]{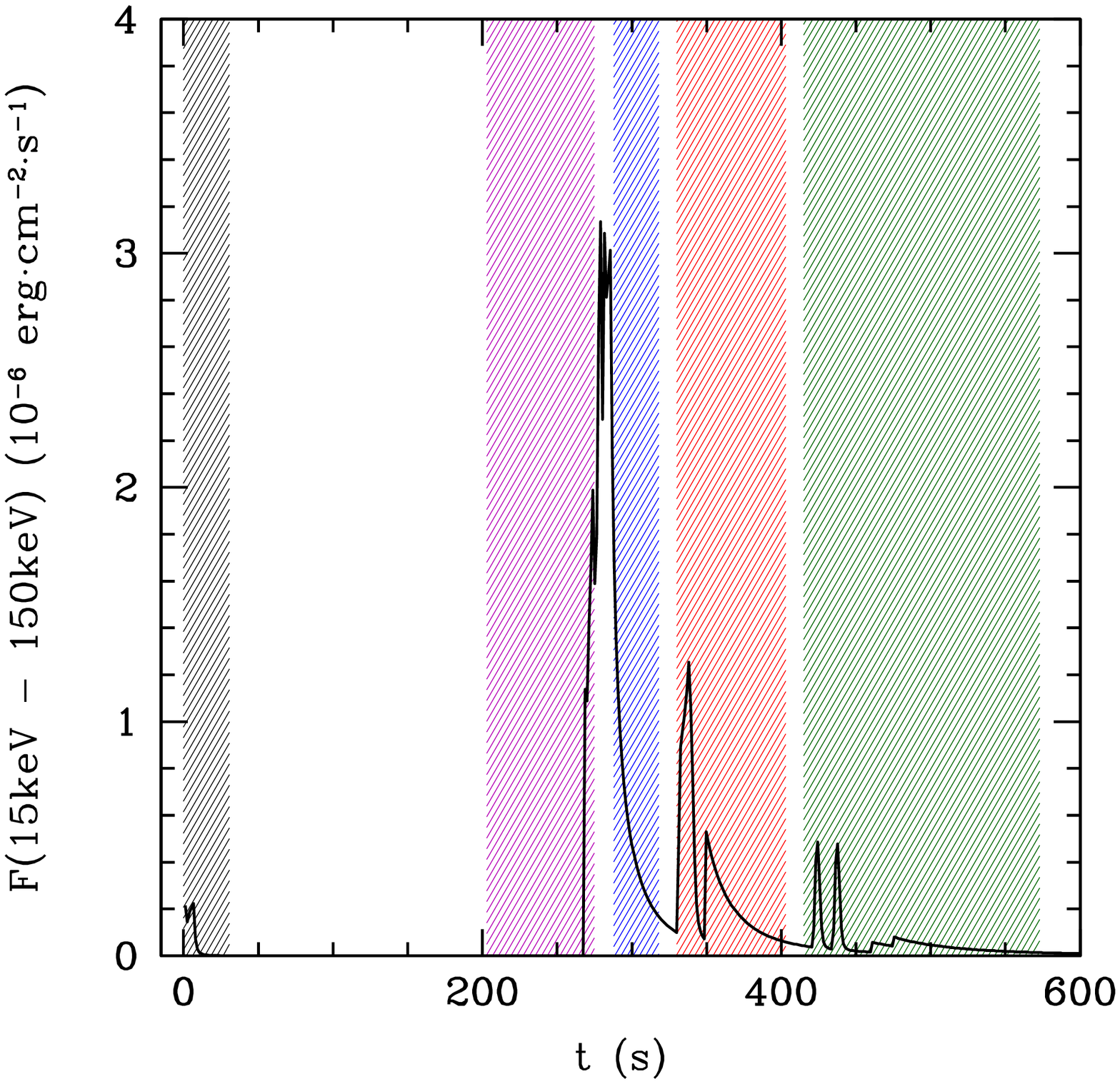} \\
\includegraphics[scale=0.32]{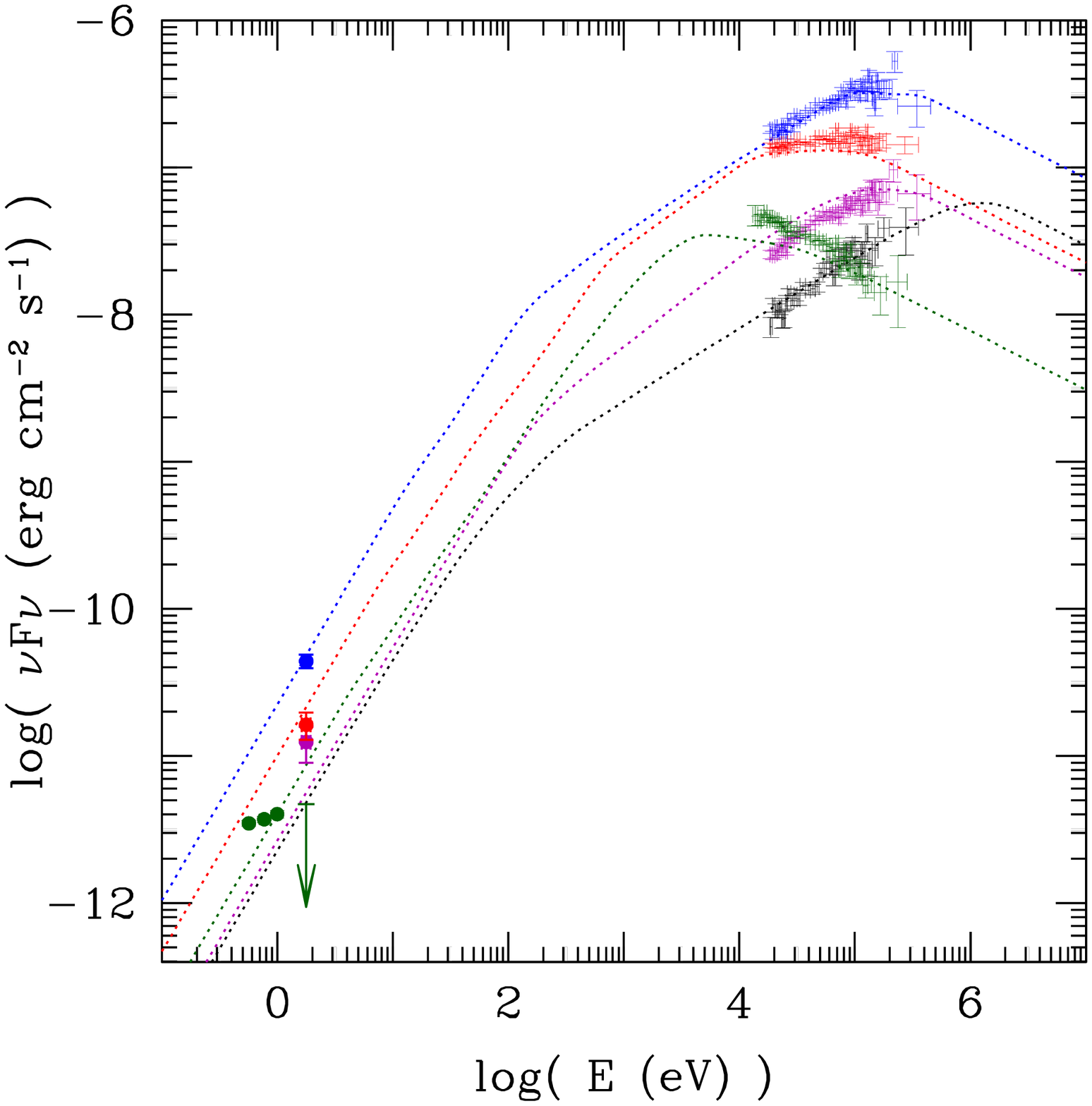}
\end{tabular}
\caption{Same as Fig.~\ref{fig:bb1} with deceleration by a wind-like external medium ($A_* = 0.1$). }
\label{fig:bb2}
\end{figure}

Hence we tried to model the temporal and spectral properties of GRB 041219A in the case where the deceleration effect of the
external medium cannot be neglected. To follow the dynamical evolution of the ejecta we then adopt the method developed by \citet{genetb2007}. The reverse and internal shocks are now contemporaneous and their interaction modifies the
observed emission: the location and intensity of the pulses in the $\gamma$-ray profile are changed, some are suppressed and others 
are created by late additional collisions in the decelerating flow. As a result, a new initial distribution of the Lorentz factor (Fig.~\ref{fig:bb2}, top panel) 
must be adopted to recover the main features of the temporal profile.      
The micro-physical parameters for the reverse shock are identical to those adopted for internal shocks. This assumption seems reasonable 
since the reverse shock, which is mildly relativistic and propagates within the ejecta, is very similar to internal shocks.
 
Satisfactory results for the temporal and spectral properties of GRB 041219A can then be obtained with a denser environment.
Fig. \ref{fig:bb2} presents an example of a possible solution for a wind like medium with $A_*=0.1$. 


\subsubsection{Infrared emission}
\label{Infrared emission}

The infrared flare detected during time interval 4 (see Table 3) cannot be easily explained by internal shocks.
Its soft spectral slope (photon index 
$\Gamma_{\rm {IR}} \sim 1.7$) 
differs from the gamma to optical average slope ($\Gamma_{\gamma - \rm{opt}}\simeq 1$) and the limited data do not allow to assess its level of correlation with the high-energy emission.

We therefore checked if it could be accounted for by a contribution from the forward shock. Then, the soft spectral slope can be 
reproduced but getting the correct time evolution appears very challenging. The late rise of the flare imposes too low values of 
the Lorentz factor and/or external density and moreover the observed variability of the infrared flux following flare maximum is not
expected from the forward shock \citep[see however][]{fan05}.  

The infrared flare may also come from a late collision in the flow but it is now 
not easy to get the right photon index. The observed $\Gamma_{\rm {IR}}$ is close to $3/2$ which would imply that $\nu_c$
is below the infrared domain whereas it was required to be above the optical at earlier times. Such a variation of
$\nu_c$ is not expected, except if the micro-physical parameters can vary by large factors
during the propagation of the outflow.

One could also try to solve the problem within a different framework, where the infrared-optical and gamma-ray components result from different radiative mechanisms. For example in the Synchro-Self Compton scenario the low-energy photons correspond to synchrotron emission from relativistic electrons while the gamma-rays result from Inverse Compton scatterings of the synchrotron photons by the same electron population. This scenario has been considered for the naked eye burst GRB 080319B \citep{racusin08,zou09,kumar08}. However it has been shown that in this case most of the dissipated energy is boosted in the GeV-TeV range by a second IC scattering (leading to an energy crisis, \citet{piran09}). Moreover if this was the common scenario for GRBs, a second peak at high energy would then be expected for most GRBs, which has not been confirmed by Fermi observations. 
Another possibility has been proposed by \citet{zheng06}. In their model the gamma-rays penetrate an electron cloud (produced in the supernova explosion prior to the GRB) and are partially converted into lower energy photons by successive Compton scatterings.

\subsubsection{Precursor activity}
As can be seen from Fig.~\ref{fig:timeres}, where the photon indices of all the spectra have been plotted versus time, there is a clear spectral evolution 
from hard to soft during the burst, superimposed on an intensity-hardness
correlation, where the hardest spectra appear in correspondence with the peaks in the light curve. 
This ``classical" evolution is however in contradiction with what is measured for the two precursors. If we take them
into account, the hardest part of the GRB is represented by the first precursor and the softest one by the second precursor.
Then the GRB becomes again very hard. 

Precursors have been subject to different definitions in the past. For instance \citet{koshut95} define precursors as
events that are dimmer than the main GRB peak and for which there is a background interval at least
as long as the main event. This does not strictly apply to \src. On the other hand, \citet{lazzati05} identifies
precursors as events that contain 0.1--1\% of the counts with respect to the main event, and are separated
by a period of quiescence without duration constraints. Our two precursors fit these requirements, since we find that they contain 1--2\% and less than 1\% of the total 
GRB flux, respectively.

Precursors of thermal origin are expected in GRBs, due to the photo-spheric emission of the fireball when
it becomes transparent \citep[e.g.][]{daigne02}, and may have been observed in the past \citep[e.g.][]{murakami91}. However in our
case neither the first nor the second precursor can be fitted with thermal spectra, and, in addition, the large delay between
the precursors and the main emission is much longer than what is expected for thermal precursors.

Different models have been invoked to explain non-thermal precursors. \citet{waxman03} and \citet{ramirez02} 
propose 
that they are associated with the jet breakout of the stellar envelope, but 
the predicted delay between the precursor and the main GRB component is again too short to be applied
to \src\ \citep{wang07}. 

More generally, the ultra relativistic motion contracts so much the durations for the observer,
that long delays are unreachable in all models where the same ejected material is supposed to radiate
at different stages of its propagation to successively produce the precursor and the prompt emission.  
Therefore considering the  
similarity between the spectral properties of the precursors and the main emission -- as also pointed out by \citet{burlon08,burlon09} using BATSE and \swi\ bursts --, 
we are left with a scenario of intermittent central engine activity, 
which remains to be explained in the context of the collapsar model.

\section{Conclusions}
We presented a detailed spectral and temporal analysis of \src\ using IBIS and SPI-ACS on board \int. We have shown that the burst presents three spectrally distinct phases (two precursors and a main event) that are separated by two long intervals of inactivity. We have analysed the available
X-ray afterglow data of \src\ constraining the column density in the direction
of this burst located at low Galactic latitude. We performed a near infrared
observation campaign of the error region of the source, using TNG and CFHT, and
identified the host galaxy of the GRB. Thanks to our multi-band photometry we
modelled the host galaxy, and classified it as an under-luminous, likely irregular galaxy at redshift $z=0.31_{-0.26}^{+0.54}$. Using this distance estimate we were able to build
the SED of the prompt emission of \src\ from the near infrared to the $\gamma$-rays.

Our modelling of the light curves and broad-band spectra obtained for GRB 041219A
shows that the prompt optical and gamma-ray emission can be explained with a common
mechanism : synchrotron radiation from shock accelerated electrons in internal
shocks. However, we point out that, due to the long duration of this burst, it is
highly probable that the deceleration of the outflow by the external medium will
start before the end of internal shocks. Our analysis shows that a more realistic
scenario taking into account this effect still agrees with the observations.
The observed gamma-ray and optical flux would then be a combination from internal
and reverse shock contributions. Our modelling has been performed with an adopted redshift $z=0.3$. Similar results would be obtained for a different z in the interval 0.05--0.85 (see section \ref{sec:identification}) with a suitable adjustment of the model parameters ($E_{\rm kin}$ and micro-physics parameters).

The late near infrared data are more
puzzling. The observed variability appears incompatible with a 
forward shock origin. On the other hand, the measured
spectral slope is difficult to reproduce by the contribution of late internal
shocks or the reverse shock, except if one assumes that the micro-physics parameters
in the outflow can vary significantly during propagation. Finally, the early episodes
in the gamma-ray light curve are not compatible with most models of precursor
activity (photospheric emission, shock breakout, etc.) due to the very long delay before
the main event. This would point out to a central engine having several distinct episodes
of relativistic ejection.

\section*{Acknowledgments}

Based on observations with INTEGRAL, an ESA project with instruments and
science data centre funded by ESA member states (especially the PI
countries: Denmark, France, Germany, Italy, Switzerland, Spain), Czech
Republic and Poland, and with the participation of Russia and the USA, and
on observations obtained with WIRCam, a joint project of CFHT, Taiwan, Korea, Canada, France, at the Canada-France-Hawaii Telescope (CFHT) which is operated by the National Research Council (NRC) of Canada, the Institute National des Sciences de l'Univers of the Centre National de la Recherche Scientifique of France, and the University of Hawaii.
ISGRI has been realized and maintained in flight by CEA-Saclay/IRFU with
the support of CNES. D.G., F.D. and R.M. acknowledge the French Space Agency (CNES) for financial support. 
R.H. is funded by the research foundation from ''Capital Fund Management''.
P.E. acknowledges financial support from the Autonomous Region of Sardinia through a research grant under the program PO Sardegna FSE 2007--2013, L.R. 7/2007 ``Promoting scientific research and innovation technology in Sardinia''.
A.F.S. acknowledges support from the Spanish MICYNN projects AYA2006-14056 and Consolider-Ingenio 2007-32022, and from the Generalitat Valenciana project Prometeo 2008/132, and the kind hospitality of the Observatori Astronomic de la Universitat de Valencia during the development of this work.
The authors are grateful to Valentina Bianchin and Luigi Foschini for providing help in PICsIT data reduction, and to the TERAPIX team ({\tt http://terapix.iap.fr/}) for providing the CHFT/WIRCam data reduction.

\bibliographystyle{mn2e}
\bibliography{biblio}



\label{lastpage}

\end{document}